\documentclass{aip-cp}
\usepackage[numbers]{natbib}
\usepackage[colorlinks,breaklinks,citecolor=blue]{hyperref}
\usepackage{rotating}
\usepackage{graphicx,epsfig,fancyhdr,psfig,epsf,txfonts,epstopdf,multirow,appendix}

\def \kms {{ \rm km\,s$^{-1}$}}

\def \arcsec {$^{''}$}
\def \siiv {Si\,{\sc IV}}

\def \ie {{i.e.\,}}

\def \A {{\AA\,}}

\graphicspath{{fig/}}
\begin{document}
\title{Transition region bright dots in active regions observed by the Interface Region Imaging Spectrograph}

\author[aff1]{Zhenyong Hou\corref{cor1}}
\author[aff1]{Zhenghua Huang}
\author[aff1]{Lidong Xia}
\author[aff1]{Bo Li}
\author[aff2]{Maria S. Madjarska}
\author[aff1]{Hui Fu}
\affil[aff1]{Shandong Provincial Key Laboratory of Optical Astronomy and Solar-Terrestrial Environment, School of Space Science and Physics, Shandong University (Weihai), 264209 Weihai, Shandong, China}
\affil[aff2]{Armagh Observatory, College Hill, Armagh BT61 9DG, UK}
\corresp[cor1]{hzyfzzf@mail.sdu.edu.cn}

\maketitle
\begin{abstract}
The Interface Region Imaging Spectrograph (IRIS) reveals numerous small-scale (sub-arcsecond) brightenings that appear as bright dots sparkling the solar transition region in active regions.
Here, we report a statistical study on these transition-region bright dots.
We use an automatic approach to identify 2742 dots in a \siiv\,raster image.
We find that the average spatial size of the dots is 0.8~arcsec$^2$ and most of them are located in the faculae area.
Their Doppler velocities obtained from the \siiv~1394~\A line range from $-$20 to 20~\kms. Among these 2742 dots, 1224 are predominantly blue-shifted and 1518 are red-shifted.
Their non-thermal velocities range from 4 to 50~\kms\ with an average of 24~\kms. We speculate that the bright dots studied here are small-scale impulsive energetic events that can heat the active region corona.
\end{abstract}

\section{INTRODUCTION}
Presently, The Interface Region Imaging Spectrograph (IRIS) observes the chromosphere and the transition region at an unprecedented high spatial resolution (0.4\arcsec). IRIS observations revealed that the solar transition region of the faculae areas of active regions are abundant of subarcsecond bright dots \citep{2014ApJ...790L..29T, 2014Sci...346D.315D, 2014Sci...346B.315T}.
In this study we statistically analyze the morphology and the plasma parameters of these features by combining IRIS spectroscopic and imaging data.

\subsection{OBSERVATIONS AND DATA ANALYSIS}
\begin{figure*}[!ht]
\centering\includegraphics[trim=3cm 2cm 2cm 2cm,scale=0.45,angle=90]{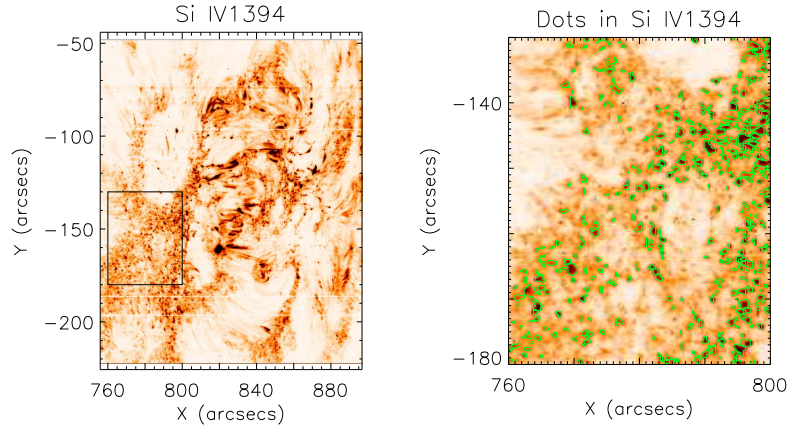}
\caption{Left: Intensity image of the IRIS \siiv\,1394\,\A raster scan. The black box outlines the FOV of the image in the right panel.
         Right: A small region taken out from the FOV  shown in the left panel.
         The contours (green) outline the bright dots.
         The images are shown in reversed colour table.}
\label{figsi}
\end{figure*}

The data used in this study are taken by IRIS\,\citep{2014SoPh..289.2733D}, from 20:19:32~UT to 20:41:47~UT on 2014 February 16.
The field-of-view (FOV) of the spectral raster scan is 141\arcsec $\times$ 175\arcsec.
The IRIS raster field-of-view is shown in the left panel of Fig.~\ref{figsi}.

We used the Southwest Automatic Magnetic Interpretation Suite\,\citep[SWAMIS,][]{2007ApJ...666..576D} feature tracking algorithm to identify the bright dots in the \siiv\,intensity image. The package is freely available (\href{http://www.boulder.swri.edu/swamis/}{http://www.boulder.swri.edu/swamis/}) and fully documented.
It was initially designed for tracking moving magnetic features on the Sun, but can also perform feature identification in any type of images, including solar EUV images.
SWAMIS contains two identification methods: ``clump'' and ``downhill''.
The clump method groups any adjacent same-sign pixels into one feature with one identification number.
The downhill method finds local maxima in the data and assigns each maximum a unique label number.
Here, we use the downhill method to find the bright dots in the raster image of the \siiv~1394~\A line.
In the right panel of Fig.~\ref{figsi}, we display the identifications of the bright dots in a small area of the active region.

The output of SWAMIS includes the spatial sizes of the bright dots and the locations of the pixels in each bright dot.
Based on this information, we performed a spectroscopic analysis of each bright dot.
In the present work, we used the \siiv~1394~\A spectral line to analyze the morphology, Doppler shifts and non-thermal velocities of the bright dots.

\section{RESULTS}
\begin{figure*}[!ht]
\centering
\includegraphics[trim=4cm 0cm 4.5cm 0cm,scale=0.53,angle=90]{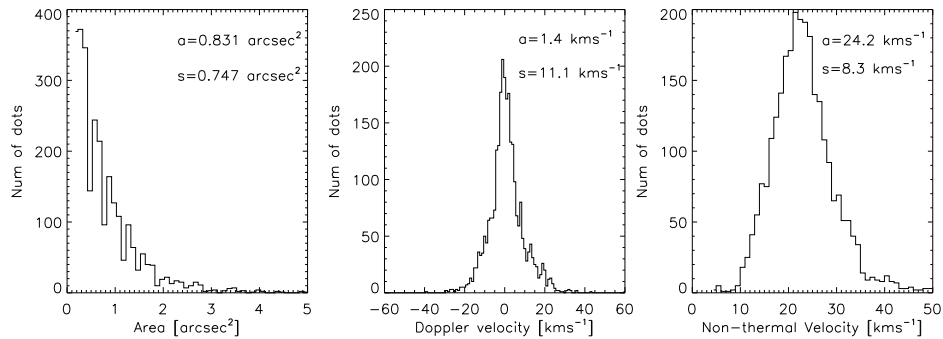}
\caption{Histograms of the physical parameters of the bright dots (Left: spatial sizes, middle: Doppler velocities, right: non-thermal velocities).
         From left panel to right: ``a'' and ``s'' give the average and the standard deviation of the area size, Doppler velocity and non-thermal velocity.}
\label{dis}
\end{figure*}

Fig.~\ref{dis} presents the histograms of the spatial sizes, Doppler velocities and non-thermal velocities of the bright dots.
The spatial sizes of the bright dots ranges from 0.2 to 16.8~arcsec$^{2}$ with an average of 0.8~arcsec$^{2}$.
The Doppler velocities of the bright dots range from $-$20 to 20~\kms, from which 1224 are predominantly blue-shifted and 1518 are red-shifted.
Their non-thermal velocities range from 4 to 50~\kms, with an average of 24~\kms.

\begin{figure*}[!ht]
\centering
\includegraphics[clip,trim=1cm 0.8cm 1cm 2cm,scale=0.8]{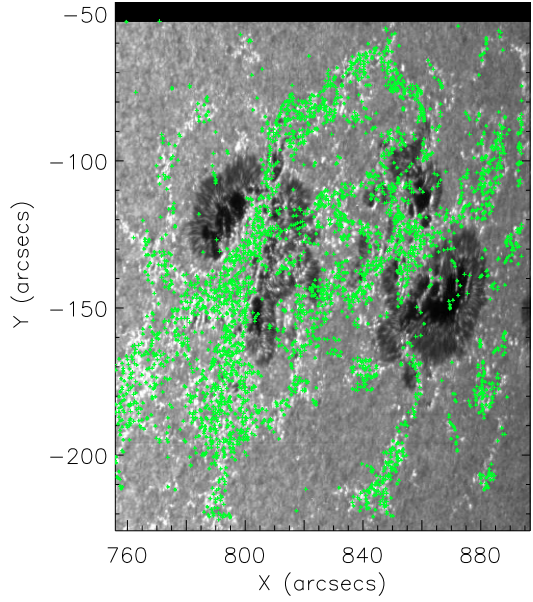}
\caption{Locations of the small-scale bright dots on the 2832~\A continuum intensity image.}
\label{loc}
\end{figure*}

We analysed the spatial distribution of the bright dots in the AR.
In Fig.~\ref{loc}, we show the intensity image of the 2832~\A continuum (chromosphere), where the locations of the bright dots are marked by green symbols. From the continuum 2832~\A intensity image, we found that about 114 dots lie in the sunspot areas (103 in the penumbra and 11 in the umbra), 1707 dots are located in the faculae areas and 921 dots in the relatively quieter areas of the AR.

\begin{figure*}[!ht]
\centering
\includegraphics[trim=2cm 8cm 2cm .5cm,scale=0.4]{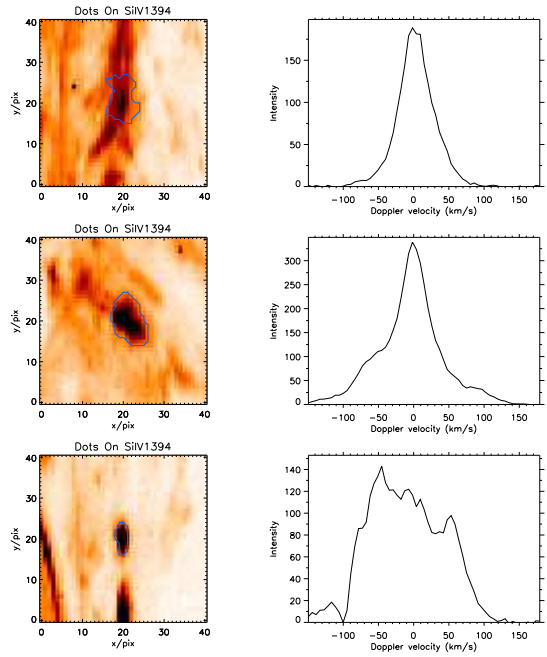}
\caption{Three typical bright dots seen in \siiv~1394~\A (left panel) and their corresponding \siiv~1394~\A spectra (right panel).
         The contours in the left panels outline the area of the bright dots identified by SWAMIS from which the average spectral line profiles shown in the right panel are obtained.}
\label{profiles}
\end{figure*}

We also studied the morphology of the bright dots \siiv\,spectra.
Three typical examples of \siiv\,profiles are shown in Fig.~\ref{profiles}.
In the top row of Fig.~\ref{profiles}, we give an example of a Gaussian line profile in a bright dot.
This type of line profiles is found in the vast majority of the identified bright dots (2432 out of 2742).
The middle row of Fig.~\ref{profiles} displays a typical transition-region explosive event line profile that is a spectrum showing enhanced wings at Doppler velocities ranging from 50 to 100~\kms \citep{1983ApJ...272..329B, 2014ApJ...797...88H}.
These line profiles are detected in 210 out of 2742 bright dots.
The remaining (100) bright dots have very broad spectral profiles also typical for explosive events \citep{1989SoPh..123...41D} with absorption-like components \citep{2014Sci...346C.315P} (see the bottom row of Fig.~\ref{profiles}).

\section{SUMMARY}

In this study, we used an automatic feature identification code, SWAMIS, to identify small-scale bright dots in \siiv\,observed by IRIS.
In total, 2742 bright dots were found, and their sizes, Doppler and non-thermal velocities as well as spectral morphologies are then statistically analyzed.

Most of the bright dots are found in the faculae area of the AR, \ie 1707 out of 2742.
From the remaining 1035, 114 are located in the sunspots (103 in penumbrae and 11 in umbra) and 921 in a relatively quiet region.
The sizes of the bright dots range from 0.2 to 16.8~arcsec$^{2}$ with an average of 0.8~arcsec$^{2}$.
The Doppler velocities of bright dots are from $-$20 to 20~\kms, from which 1224 are blue-shifted and 1518 are red-shifted.
Their non-thermal velocities range from 4 to 50~\kms, with an average of 24~\kms.

\par
The \siiv\,spectra of most of the bright dots (2432 out of 2742) are Gaussian.
A very few (310) bright dots show non-Gaussian spectra, from which 210 have a typical explosive-event profile and 100 display very broad profiles with absorption-like components.

\par
The bright dots analysed in this study are likely to be associated with nano-flares or other fine structures, such as EUV bright dots \citep[EBDs,][]{2014ApJ...784..134R}, hot bombs \citep{2014Sci...346C.315P}, sunspot bright dots \citep{2014ApJ...790L..29T} and transition region explosive events \citep{1983ApJ...272..329B, 2014ApJ...797...88H}. All these events have been suggested by those authors to be signatures of energy release in the solar atmosphere. We speculate that the bright dots studied here are small-scale impulsive energetic events that can heat the active region corona. This speculation will be investigated in our following-up study.

\section{ACKNOWLEDGMENTS}
This research is supported by the China 973 program 2012CB825601, and the National Natural Science Foundation of China under contract 41274178, 41404135, 41474150.
IRIS is a NASA small explorer mission developed and operated by LMSAL with mission operations executed at NASA Ames Research center and major contributions to downlink communications funded by the Norwegian Space Center (NSC, Norway) through an ESA PRODEX contract.
AIA and HMI data are courtesy of SDO (NASA).
The SWAMIS was written by Craig DeForest and Derek Lamb at the Southwest Research Institute Department of Space Studies in Boulder, Colorado.
MM is funded by the Leverhulme trust, UK.
\nocite{*}
\bibliographystyle{aipnum-cp}%
\bibliography{bibliography2}
\end{document}